# Electrically pumped single-defect light emitters in WSe$_2$

S. Schwarz[1*], A. Kozikov[2*], F. Withers[2], J. K. Maguire[1], A. P. Foster[1], S. Dufferwiel[1], L. Hague[2], M. N. Makhonin[1], L. R. Wilson[1], A . K. Geim[2], K. S. Novoselov[2], A. I. Tartakovskii[1]

[1]Department of Physics and Astronomy, University of Sheffield, Sheffield S3 7RH, UK
[2]School of Physics and Astronomy, University of Manchester, Manchester M13 9PL, UK
* These authors contributed equally

**Recent developments in fabrication of van der Waals heterostructures enable new type of devices assembled by stacking atomically thin layers of two-dimensional materials. Using this approach, we fabricate light-emitting devices based on a monolayer WSe$_2$, and also comprising boron nitride tunnelling barriers and graphene electrodes, and observe sharp luminescence spectra from individual defects in WSe$_2$ under both optical and electrical excitation. This paves the way towards the realization of electrically-pumped quantum emitters in atomically thin semiconductors. In addition we demonstrate tuning by more than 1 meV of the emission energy of the defect luminescence by applying a vertical electric field. This provides an estimate of the permanent electric dipole created by the corresponding electron-hole pair. The light-emitting devices investigated in our work can be assembled on a variety of substrates enabling a route to integration of electrically pumped single quantum emitters with existing technologies in nano-photonics and optoelectronics.**

The recent observation of direct bandgaps in semiconducting molybdenum and tungsten dichalcogenide monolayers has led to a rise of interest to these two-dimensional (2D) materials and demonstrated their potential for future optoelectronic devices[1,2,3,4]. These one-monolayer-thick crystals are characterised by large exciton binding energies[5,6] and oscillator strengths[7] and can be combined with other layered materials to create heterostructures held together by van der Waals forces[8,9,10,11,12]. This concept has been used to form electrically driven light-emitting structures, where MoX$_2$ or WX$_2$ (X=S or Se) monolayers were used as the exciton recombination layers, thin hexagonal boron nitride (hBN) was used for tunnelling barriers and graphene - used for transparent electrodes[11,12].

In these devices, in addition to the pronounced 2D excitonic peaks that could be excited both optically and electrically, various (usually broad) features have also been detected 50 to 150 meV below the neutral exciton peak in both photo- and electroluminescence[1,2,13,14,15,16,17,10,11]. These were assigned to luminescence of localized states associated with defects in the monolayer semiconductors. In this energy range it was also possible to observe narrow PL features with linewidths 100-400 μeV particularly pronounced in WSe$_2$ at temperatures below 30 K[14,15,16,17,18,19,20]. The origin of the defect states giving rise to such PL response remains unknown. However, it has been shown for WSe$_2$ that photons emitted into the sharp defect PL peaks exhibit anti-bunching[14,15,16,17,18,19]. This provides a unique single-photon source contained within an atomic monolayer of a solid state material, which opens the way for integration of such defect light emitters with various photonic structures such as photonic crystal[21], microdisk[22] and scanning cavities[23] as well as waveguides.

Here we demonstrate electrical pumping of such defect states in graphene/hBN/WSe$_2$/hBN/graphene heterostructures previously used to generate electroluminescence (EL) from 2D excitons. We identify the regimes where EL of single defects dominates the emission spectrum of the device as the filling of the localized states occurs at a lower bias voltage ($V_b$) than for the 2D states. Overall, EL appears at applied bias around -2.1 V, close to the expected single-particle bandgap in $WSe_2$. We also show that our device can be used to fine-tune the energy of the defect emitter by varying the bias, constituting observation of the quantum confined Stark effect[24,25].

The upper part of Fig.1(a) shows a schematic of the van der Waals heterostructure that forms the EL device. Here, the graphene layers serve as transparent electric contacts and the hBN layers are used as tunnelling barriers that separate the graphene contacts from the active material, a $WSe_2$ monolayer. The lower part of Fig.1(a) shows a microscope image of one of the heterostructures used in this work, where graphene contacts and a $WSe_2$ monolayer are clearly visible. In this image the hBN layers have a low contrast and are invisible, whereas they can be imaged using various optical filter arrangements and dark field microscopy.

A schematic of the band structure for a finite $V_b$ applied between the graphene contacts and corresponding to the onset of EL is shown in Fig.1(b). The quasi-Fermi levels in the graphene contacts shift with increasing voltage. As the quasi-Fermi level of the left contact reaches the minimum of the conduction band of $WSe_2$, electrons tunnel through the hBN barrier. Holes tunnel through the other hBN barrier (on the right) when the quasi-Fermi level of the corresponding contact reaches the maximum of the valence band of $WSe_2$. Upon injection of both electrons and holes into $WSe_2$ layer the formation of excitons becomes possible. A typical IV-curve for this EL device is shown in Fig.1(c), where the EL onset voltage is observed at the 'kink' in the curve for a bias of -1.9V.

In addition, electron-hole pair excitation can be achieved in a wide voltage range using optical pumping above the bandgap, which creates excitons and electron-hole pairs relaxing into the lowest exciton states and giving rise to PL. In our experiments we use a typical micro-PL arrangement, where for PL measurements a laser is focussed in a 2 μm diameter spot on the sample by a short focal length lens positioned above the sample in a gas-exchange cryostat at a temperature of 4.2K. The lens has a high numerical aperture of 0.55 enabling efficient collection of PL and EL from the sample. The PL and EL are then sent through a single-mode fibre to a spectrometer and measured with a charge coupled device. The single-mode fibre effectively acts as a spatial filter collecting light from a 2.5 μm diameter spot on the sample. In PL measurements described below we used a laser with photon energy 1.94 eV and the laser power was kept at 300 (30) μW for the data in Fig.1 (Fig.2).

A typical PL spectrum of a monolayer $WSe_2$ is shown in Fig.1(d). The PL features $X^0$ and $X^-$ are attributed to neutral exciton and trion peaks, whereas the features P0-P3 correspond to PL from the localized exciton states. Previously single photon emitters have been found in the whole range of energies where we observe the features P0-P3[14,15,16,17,18,19]. The PL bias dependence is shown in Fig.1(e). For biases between -1.5 V and -1.9 V the quasi-Fermi level of the top graphene layer reaches the minimum of the conduction band resulting in supply of additional electrons. This leads to an increase of the trion contribution in PL just before the onset of EL. At $V_b$<-1.9 V, the $X^-$ intensity drops, but overall the sample luminescence becomes significantly brighter with particular enhancement at lower energies below 1.67 eV. This corresponds to a situation when the tunnelling of both electrons and holes is possible from the graphene contacts. In this regime, the luminescence in the sample is excited mainly through electrical injection. In order to compare with typical PL, a an EL spectrum in this regime is shown in Fig.1(f). The main differences are in the lower intensity of both $X^0$ and $X^-$ relative to very pronounced low energy peaks, with the strongest intensity observed for P1. Similar behaviour is observed for positive bias.

At a lower laser power of 30 μW in some locations on the sample it is also possible to isolate narrow PL peaks having linewidths down to 100 μeV. Fig.2(a) shows an example of such peaks in the energy range corresponding to the localized state PL. Similarly to the observation in the III-V quantum dots[26], pairs of orthogonally-polarised PL lines are typically observed. The splitting between the lines is typically about 1 meV, with the low energy component of the doublet being more pronounced. Fig.2(b) shows that the doublet has a strong linear cross-polarization with a polarization degree of 90%. The observed splitting is possibly a consequence of a low symmetry of the defect, whereas the difference in intensity may be explained by relaxation with acoustic phonon emission efficient at low temperature[27].

By varying the bias $V_b$ steady shifts of some defect PL lines have been observed as shown in Figs.2(c,d). The ability to spectrally tune a quantum emitter is an essential feature for potential quantum interference experiments[28]. However, in our case a range of different behaviours is observed with different peak energy shift rates, appearance and disappearance of new lines in narrow voltage ranges etc. Fig.2(d) shows this behaviour for a bright PL peak at 1.632 eV. In addition to the energy shift with a rate of about 0.4 meV/V, fluctuations in the photon energy are observed including discontinuities around ±1.5 and 0 V. This may be related to a long-term recharging of the defects in the vicinity of this light-emitting centre.

From the trend in the peak position we can estimate the permanent dipole formed by the electron-hole pair localized on the defect [Fig.2(d)]. The change in the emission energy is given by $\Delta U = -p\Delta V/d$, where $p$ is the electric dipole, $\Delta V$ is the change in the applied voltage and $d$ is the distance between the graphene contacts. We obtain that $p/ed=0.4 \times 10^{-3}$, which for $d \approx 3$ nm gives $p/e \approx 1.2$ pm indicating a vanishing vertical displacement of the electron and hole wavefunctions. This estimation has been enabled by a well-understood distribution of the electric field in our structure, where the most significant contribution is expected to be from the vertical electric field.

We will now discuss the results of the EL measurements on a single-defect emitter (SDE). The panels of Fig.3(a) show typical EL spectra observed for a defect emitter for different applied voltages (no laser excitation is present here). Surprisingly, there is a regime where the signal from the single-defect emitter (SDE) dominates the EL spectrum as shown in the bottom panel for $V_b$=-2.15 V, where a single sharp line labelled SDE1 is observed at 1.607 eV. As the bias is increased, additional broad features first appear at energies around the emitter, where luminescence from localized states is expected [e.g. P3 in Fig.1(c)]. For lower biases other features become dominant with most prominent peaks P1 and P2. At around -2.3 V, trion EL also becomes visible around 1.7 eV. The peak intensity of the SDE1 line remains high (much stronger than for the $X^-$ line) and still comparable with the dominant features P1 and P2 up to the bias of -2.5 V. We find that EL from SDE1 peak and other SDE lines is localised in an area on our sample comparable with the spatial resolution provided by the single-mode fibre in the light collection path, in agreement with the assumption that this signal comes from a single localized defect site.

The purity of a single-photon emitter (the depth of the anti-bunching curve) depends on how efficiently other luminescence can be suppressed, i.e. ideally a spectrally isolated emitter is required[29]. Although in our case the SDE1 EL is superimposed on a broad background of EL from other states, there are bias regimes where its EL is the strongest. In order to demonstrate this further, in Fig.3(b) we show bias dependences of the normalized integrated intensity for the SDE1 peak, EL of the localized states (integrated up to 1.635 eV) constituting the background around the SDE1 peak, and EL above 1.635 eV including P2, P1 and $X^-$. Up to about -2.15 V both the localized states emission and the high energy features are not observed in EL, whereas the SDE peak intensity is already considerable. Below -2.2 V, the high energy EL grows fast with bias, whereas the SDE EL gradually saturates and then decreases for $V_b$<-2.35V.

Fig.3(c) shows a typical bias dependence of the full-width-at-half-maximum (FWHM) of the EL peak for SDE1 peak, with a minimum of 0.6 meV observed at low biases where EL emerges and a subsequent increase in the saturation regime above -2.3 V to 1.4 meV. Typically SDE EL has an asymmetric shape, with a pronounced sharp maximum mostly contributing to the FWHM [see Fig.3(d)]. The main maximum in the SDE spectra is also accompanied with a high energy shoulder about 1 meV above and with a more pronounced low energy shoulder 2-2.5 meV below. While the high energy shoulder may be caused by the fine-structure splitting observed in Fig.2(a), the origin of the low energy shoulder is not known. The low linewidth of the main maximum may allow efficient spectral filtering to improve the purity of the single photon source based on the SDEs. To demonstrate the potential of this approach, in Fig.3(d) we show an SDE EL spectrum that is filtered using a 2nm bandpass filter, which is sufficient to remove most of the background EL.

In conclusion, both photo- and electroluminescence from spatially and spectrally isolated single-defect emitters has been observed in van der Waals heterostructures using $WSe_2$ as the optically active material. In the EL regime single-defect emitters with linewidths of around 0.6 meV are most efficiently pumped at low biases close to the onset of the EL in the whole structure, where their luminescence is observed with little background. At higher biases, presence of emission from other localized states leads to a stronger background in EL, which would result in the reduced anti-bunching contrast. This effect could be partially suppressed by spectral filtering before light reaches the detection system. We also observe tuning of the single-defect emitter frequency by applied bias, an important feature in quantum interference experiments. This tuning enables estimation of the permanent electric dipole $p/e \approx 1.2$ pm of the electron-hole pair associated with the defect emission.

At the last stages of preparation of this manuscript we have become aware of two papers where similar electrical pumping has been observed in TMDC/graphene heterostructures of a different design[30,31].

**Acknowledgements**

This work was supported by the Graphene Flagship (Contract No. NECTICT-604391), the EPSRC Programme Grant EP/J007544/1 and grant EP/M012727/1. FW acknowledge support from Royal Academy of Engineering. AKG and KSN also acknowledge support from ERC, EPSRC (Towards Engineering Grand Challenges and Fellowship programs), the Royal Society, US Army Research Office, US Navy Research Office, US Airforce Research Office. AKG was supported by Lloyd's Register Foundation.

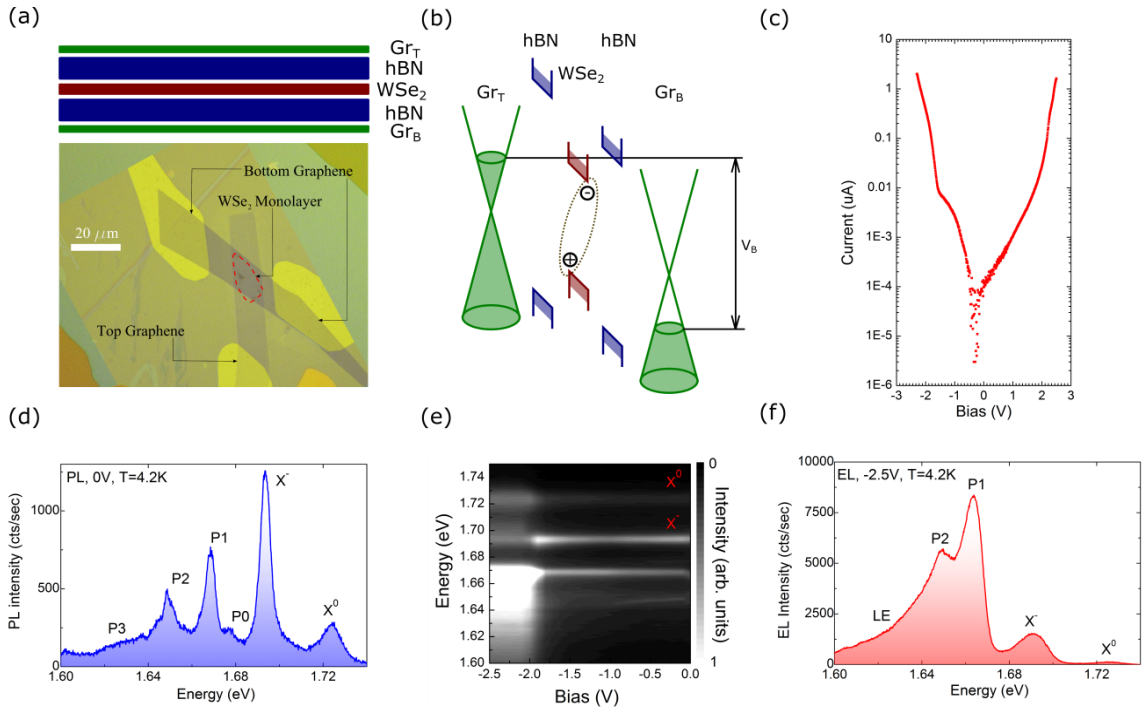

**Figure 1:** (a) Schematic of the van der Waals heterostructure used to form an electrically pumped light-emitting device. The lower panel shows an optical microscope image of the device. (b) Schematic of the band-structure of the device in (a) demonstrating electrical injection of carriers through hBN layers. (c) Typical $IV_b$ dependence (T=4.2K) of the device showing the EL onset voltage at -1.9V. (d) Typical PL spectrum of a monolayer $WSe_2$ ($T$=4.2K and $V_b$=0V) showing neutral exciton ($X^0$) and trion ($X^-$) features as well as localized peaks P0-P3. A laser with photon energy 1.94 eV and the power 0.3 mW was used. (e) PL dependence on bias voltage (T=4.2K). (f) Typical EL spectrum with no laser excitation at $T$=4.2K and $V_b$=-2.5V.

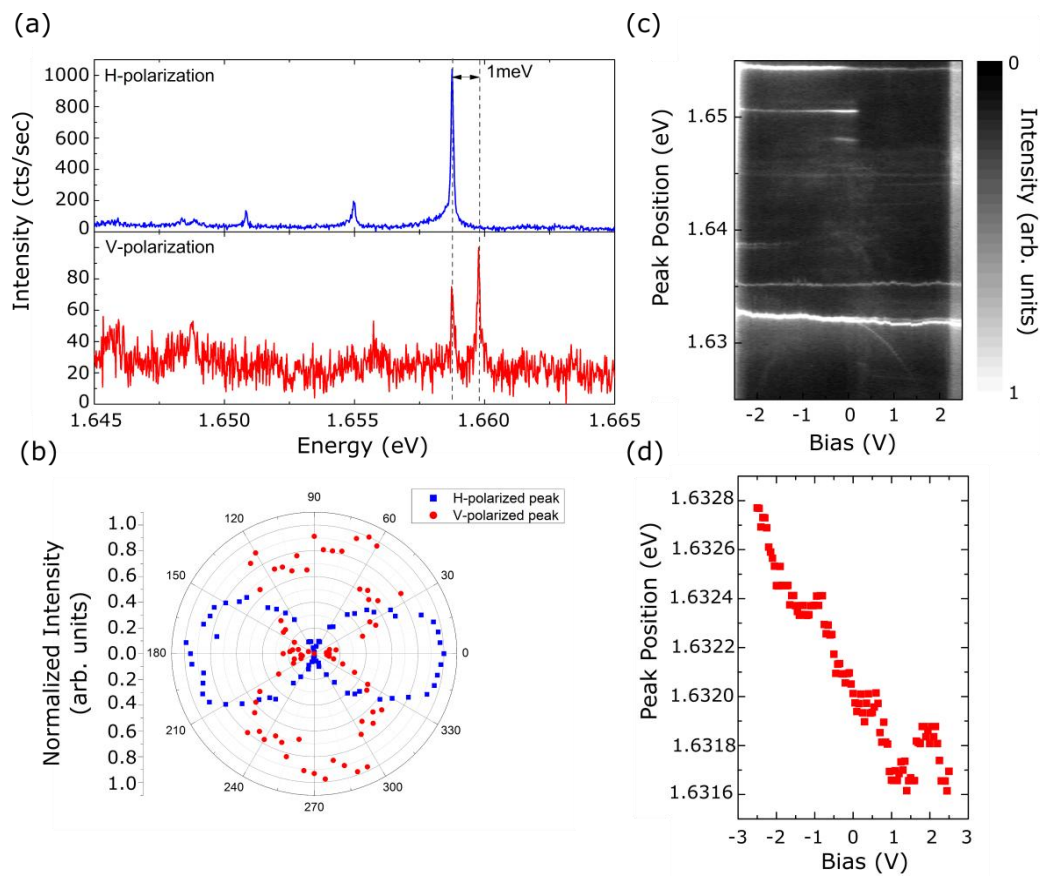

**Figure 2:** (a) Polarization resolved PL spectra of isolated defect light emitter. The linewidth of the quantum emitters is around 100 µeV with a fine-structure splitting of 1 meV. (b) Normalized PL intensity as a function of the angle of the transmission axis of the linear polarizer for the predominantly H- and V-polarized peaks. (c) PL spectra measured in the area of the sample showing single defect peaks as a function of bias voltage, $V_b$. (d) Energy position of one of the peaks in (c) as a function of $V_b$. A Stark shift of 0.4 meV/V is observed.

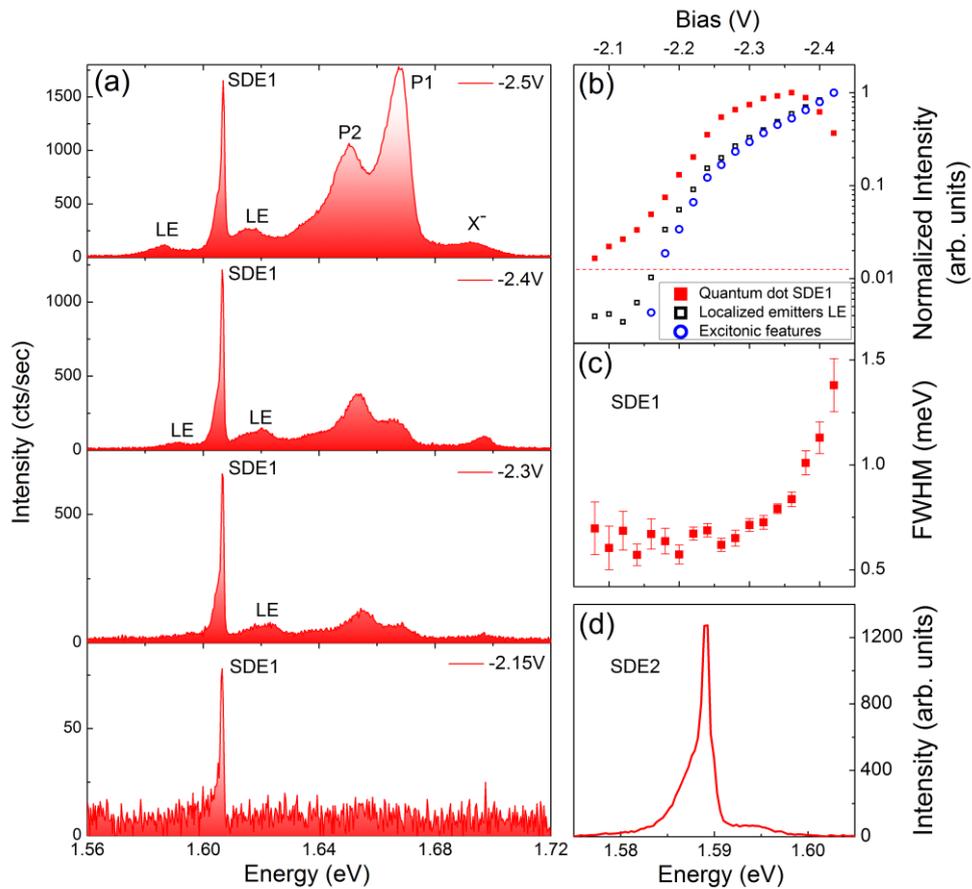

**Figure 3:** (a) Evolution of EL spectra with bias. Labels SDE1 and LE are used to denote the EL peaks of the single-defect emitters and other localized states, respectively. (b) Normalised integrated EL intensity of the SDE1, localized states (LE) and $X^-$ peaks as a function of bias voltage. The red dashed line indicates the noise floor for SDE1. (c) SDE1 peak linewidth as a function of bias voltage. (d) Spectrally filtered EL from SDE2 obtained using a 2nm bandpass filter.